# Fragmentation cross-sections and binding energies of neutron-rich nuclei

M. B. Tsang,[1] W. G. Lynch,[1] W. A. Friedman,[2] M. Mocko,[1] Z.Y. Sun,[3] N. Aoi,[4] J. M. Cook,[1] F. Delaunay,[1] M. A. Famiano,[1] H. Hui,[1] N. Imai,[4] H. Iwasaki,[5] T. Motobayashi,[4] M. Niikura,[6] T. Onishi,[5] A. M. Rogers,[1] H. Sakurai,[5] H. Suzuki,[5] E. Takeshita,[7] S. Takeuchi,[4] and M. S. Wallace[1]

[1] *National Superconducting Cyclotron Laboratory and Department of Physics & Astronomy, Michigan State University, East Lansing, Michigan 48824, USA*

[2] *Department of Physics, University of Wisconsin, Madison, Wisconsin 53706, USA*

[3] *Institute of Modern Physics, CAS, Lanzhou 730000, China*

[4] *RIKEN Nishina Center, 2-1 Hirosawa, Wako, Saitama 351-0198 Japan*

[5] *Department of Physics, University of Tokyo, 7-3-1 Hongo, Bunkyo, Tokyo 113-0033, Japan*

[6] *Center for Nuclear Study, University of Tokyo (CNS) RIKEN Campus, 2-1 Hirosawa, Wako, Saitama 351-0198, Japan*

[6] *Rikkyo University, 3 Nishi-Ikebukuro, Toshima, Tokyo 171, Japan*

## Abstract

An exponential dependence of the fragmentation cross-section on the average binding energy is observed and reproduced with a statistical model. The observed functional dependence is robust and allows the extraction of binding energies from measured cross-sections. From the systematics of $^{75,77,78,79}$Cu isotope cross-sections have been extracted. They are 636.94±0.40 MeV, 647.1±0.4 MeV, 651.6±0.4 MeV and 657.8±0.5 MeV, respectively. Specifically, the uncertainty of the binding energy of $^{75}$Cu is reduced from 980 keV (listed value in the 2003 mass table of Audi and Wapstra) to 400 keV. The predicted cross-sections of two near drip-line nuclei, $^{39}$Na and $^{40}$Mg, from the fragmentation of $^{48}$Ca are discussed.

Mapping the nuclear landscape boundaries has been a major thrust of nuclear physics research with radioactive beams [1,2]. The neutron drip-line, which marks the boundary between particle stable and unstable nuclei, is often discussed in terms of its dependence on the nuclear charge. Since the maximum neutron number for a given element is strongly influenced by the pairing interaction, it may be more reasonable to define the neutron drip-line as the lightest particle-stable isotone for each neutron number, N [2]. In this definition, the neutron drip-line may already be determined for odd neutron number nuclei up to N=27, but not determined for nuclei with N=10, 14, 22, 24, 26, and 28 [2].

The development of more intense $^{48}$Ca (Z=20, N=28) beams at present facilities should achieve the goal of determining drip-line nuclei with neutron numbers N=24-28, and possibly as a function of Z up to Z=11 [3]. This implies establishing the existence or non-existence of certain nuclei. The cross-section that establishes the existence of a new nucleus places constraints on the masses of nuclei in the neighborhood of the neutron drip-line and on the effective interactions and nuclear structure models used to predict them [4]. This paper shows how the observed exponential dependence of the cross-sections on average binding energies can be used to extract binding energy. The systematics can also be used to predict the cross-sections of unknown nuclei, which is useful in planning experiments to determine their existence or non-existence.

Projectile fragmentation has been one of the preferred reaction mechanisms to produce nuclei near the drip-lines [2, 5]. This mechanism has a complex time dependence that begins with the deposition of energy into projectile spectator nuclei and ends with the sequential decay of excited projectile residues into particle stable nuclei. In order to understand this mechanism, we have measured comprehensive cross-sections of 10 projectile fragmentation reactions, 140 MeV per nucleon $^{40}$Ca, $^{48}$Ca, $^{58}$Ni, $^{64}$Ni and 64 MeV per nucleon $^{86}$Kr projectiles on $^{9}$Be and $^{181}$Ta targets [6, 7, 8].

Different reaction models, including the widely used Abrasion-Ablation model [9, 10, 11], the hybrid statistical dynamical Heavy Ion Phase Space Exploration (HIPSE) Model [12], and the Asymmetrized Molecular Dynamic Models (AMD), [13] have been compared to these data [7,14]. Each of these models produces excited projectile residues. A two-stage approach was employed that allowed these excited residues to decay sequentially. These calculations could describe some aspects of the most probable residues near the valley of stability [7,14], but all calculations under-predicted the rarest residues near the drip-lines. In most cases, the final distribution of fragments seemed to reflect more strongly the phase space considerations of the sequential decay in the second stage of the reaction than of the earlier dynamical stage.

In this paper, we assume that phase space plays a dominant role in the fragment production mechanism and describe the fragment yields within a statistical approach. We note that projectile residues are in diffusive contact with "participant" nucleons in the overlap region of projectile and target, and that the relative isotopic yields have been modeled thermally to extract isotope temperatures [15]. This suggests that there may be a practical utility to such phase-space approaches even in cases where equilibrium may not be achieved.

For simplicity we fit the fragment cross-sections with an expression that can be derived using the grand canonical ensemble. In this approach, the yield of a fragment with N neutrons and Z protons can be written as:

$$Y(Z, N) = cA^{3/2}\exp[(N\mu_n + Z\mu_p - F)/T] \qquad (1)$$

where c is the normalization constant, the mass number A=N+Z, $\mu_n$ and $\mu_p$ are the neutron and proton chemical potentials, respectively, T is the temperature and F=F(Z,N,T) is the Helmholtz free energy. F can be decomposed into a ground state contribution and the free excitation energy, $F = E_o + F^*$. The ground state energy is given by the binding energy, $E_o = -B(Z,N)$. Following [16], we approximate $F^*$ by considering particle bound levels only. Particle unbound levels decay to the particle bound levels. We assume that the final yields after decay are proportional to the particle stable levels, which are primarily located at energies below the minimum of the separation energies for protons, neutrons and alpha particles. If these bound levels are approximated by a back-shifted Fermi gas level density, then $F^*(Z,N,T) \approx -E_{FG}^*(Z,N,T) + F_0(Z,N)$ where the pairing interaction dominates $F_0(Z,N)$. For simplicity in our bound level approximation, $E_{FG}^*$ is bounded by the continuum threshold. We define $F^* \approx -0.5 \cdot \min(S_n, S_p, S_\alpha) + F_0(Z,N)$, where $S_n$, $S_p$, and $S_\alpha$ are the neutron, proton and alpha separation energies. $F_0(Z,N) = \frac{1}{2}[(-1)^N + (-1)^Z] F_{p0} \cdot A^{-3/4}$ is the ground state pairing energy. The free parameters c, $\mu_N$, $\mu_p$, $F_{p0}$ and T can be constrained by cross-section data.

We illustrate this approach using the fragment cross-sections measured in the projectile fragmentation of $^{86}$Kr+$^9$Be at E/A = 64 MeV. This reaction produces wide distributions of nickel and copper isotopes extending out to N=50 in the case of copper. We obtained the best fit values of c=2.5x10$^{-8}$ mb, $\mu_n$=-9.5 MeV, $\mu_p$=-7.0 MeV, $F_{p0}$=47 MeV and T=2.2 MeV using the cross-sections of nickel isotopes with A=64-72 and copper isotopes with A=61-79. The cross-sections of the measured copper isotopes are plotted in Figure 1 (solid symbols) as a function of mass number A. The model fit is shown as the dashed line.

Empirically, it has been observed that yields of neutron-rich isotopes within an element depend exponentially on the average binding energy per nucleon, <B>=B/A [17]. The latter observation is best illustrated in Figure 2 by plotting the experimental cross-sections of the $^{68-79}$Cu (Z=29) isotopes as a function of <B'>=(B-$\varepsilon_p$)/A, where $\varepsilon_p = \frac{1}{2}[(-1)^N+(-1)^Z]\varepsilon \cdot A^{-3/4}$ minimizes the observed odd-even variations in the cross-section. The solid line in the figure is the best exponential fit of the empirical relation:

$$\sigma = C\exp(<B'>/\tau) \qquad (2)$$

with the inverse slope, $\tau$=0.0213MeV, $\varepsilon$=6 MeV and C=2.17x10$^{-15}$.

Within the statistical approach we outline above, Equation (2) cannot be obtained as an approximation to Equation (1). It is therefore surprising to observe in Figure 2 that Eq. (1) also predicts a nearly exponential dependence on <B'> as shown by the dashed lines in the figure. Both fits roughly follow the trend of the isotopic distribution towards the lighter masses, even to the region of the peak near $^{63}$Cu as shown in Figure 1.

The correlation between mass and cross-sections in both Equations (1) and (2) can be used to determine the binding energy and its uncertainty from the measured cross-section. From Eq. (2), the binding energy uncertainty is related to the uncertainties in the cross-section measurements by

$$dB \sim T \cdot (d\sigma/\sigma) \quad (3)$$

If T~2 MeV, a 15% cross-section measurement, which should be achievable, would mean a binding energy uncertainty of about 300 keV.

In recent measurements with the ISOLTRAP, the masses of $^{65-74}$Cu and $^{76}$Cu isotopes have been accurately determined [18]. However due to technical difficulties, the mass of $^{75}$Cu was not measured. From the cross-section measurement of $^{75}$Cu we obtain (2.56±0.61) x $10^{-5}$ mb, which corresponds to the binding energy of $^{75}$Cu to be 636.94±0.40 MeV. This value is more precise than the value of 636.75 ± 0.98 MeV given in ref [19]. No direct mass measurements have been made beyond $^{76}$Cu. Our extrapolated values for the binding energies of $^{77,78,79}$Cu are 647.1±0.4 MeV, 651.6±0.4 MeV and 657.8±0.5 MeV, respectively. Within experimental uncertainties, these values are consistent with the values listed in ref. [19]. Even though the uncertainties do not represent improvement in the listed uncertainties in the extrapolated binding energy values given in ref. [19], the extracted binding energies presented here are measured values. Moreover, this simple technique of extracting masses from cross-sections could be applied generally. Until more accurate mass measurements using traps are available, improvement of the binding energies for $^{77-80}$Cu, and $^{75}$Cu nuclei, to 200-300 keV can be achieved by measuring the corresponding cross-sections more accurately.

One of the virtues of fitting the isotopic yield is the possibility to extrapolate the measured cross-sections to unmeasured isotopes. To illustrate such an extrapolation, we limited the range of copper masses in the fit with Eq. (1) to A=61-75 and extrapolated the fitting function obtained to A=79. The result (thin solid line) cannot be distinguished from the dashed line in Figure 1.

In the near future, the development of more intense $^{48}$Ca beams at present facilities should achieve the goal of determining of drip-line nuclei up to Z=11 [3]. Establishing the existence of $^{40}$Mg and non-existence of $^{39}$Na are essential for such quest. Figure 3 shows the cross-section systematics for $^{48}$Ca+$^9$Be (left panel) and $^{48}$Ca+$^{181}$Ta (right panel) reactions of neutron-rich Na and Mg isotopes with average binding energy values of less than 8 MeV. The solid lines are the extrapolation to all the heavier Na and Mg isotopes with the binding energy given in ref [19] using the parameters of ref. [17]. For Mg isotopes, the lines end at $^{40}$Mg giving the predicted $^{40}$Mg cross-sections to be 1-2x10$^{-11}$ mb for the $^{48}$Ca+$^9$Be reaction and 4-8x10$^{-11}$ mb for $^{48}$Ca+$^{181}$Ta reaction. Fitting the measured cross-section data of $^{23-31}$Na and $^{26-35}$Mg with the model parameters of Eq. (1) (solid lines) yields the predictions of $^{40}$Mg cross-sections to be 2 x10$^{-11}$ mb for the $^{48}$Ca+$^9$Be and 4 x10$^{-11}$ mb for $^{48}$Ca+$^{181}$Ta reactions. Both fits with empirical equations and the model of Equation 1 give similar results.

If the existence of $^{40}$Mg can be confirmed, an accurate cross-section for this nucleus may provide an estimate of the binding energy of $^{40}$Mg. In that case, the non-existence of $^{39}$Na would determine the neutron drip-line of N=28. The heaviest known Na isotope is $^{37}$Na [20]. $^{38}$Na is unbound. If $^{39}$Na is particle bound, it must have a larger binding energy than that of $^{37}$Na, 234.77±0.96 MeV. This means that the average binding energy of $^{39}$Na will be greater than 6.02 MeV, corresponding to a lower limit on the $^{39}$Na cross-section of (1-2)x10$^{-11}$ mb and (6-12) x10$^{-11}$ mb for Be and Ta targets respectively, assuming the exponential fit of Eq. (2) to be valid. However, the fit using the model of Eq. (1) does not predict a purely exponential dependence of the cross-sections with average binding energy. Instead, there is a downward curvature for the predicted cross-section leading to a much lower estimate for the $^{39}$Na cross-section of 5.32x10$^{-13}$ and 3.46x10$^{-12}$ mb for $^{48}$Ca+$^{9}$Be and $^{48}$Ca+$^{181}$Ta reactions, respectively. The differences between the two extrapolations are small for nuclei whose masses are only a couple nucleons heavier than the measured values, but become larger for heavier nuclei. To obtain better extrapolations, the fragmentation cross-sections of neutron-rich $^{32-35}$Na need to be measured accurately.

At present, the justification for the exponential dependence of the cross-sections on average binding energy given by Eq. (2) is purely empirical. Fits using Eq. (1) based on statistical model support this exponential dependence. In the model the decrease of the cross-section occurs through a competition between the total binding energy, which increases the cross-sections with A, and the neutron chemical potential, which decreases the cross-sections with N. At sufficiently large N, the phase space and neutron number constraints that give rise to the neutron chemical potential must require higher-order terms that will further reduce the cross-sections for such nuclei below these estimates.

In summary, we have shown that the exponential dependence of the fragment cross-sections to the average binding energy can be described by a statistical model that takes into account the constraints from phase space and conservation laws. The relation between average binding energies and the cross-sections allows the determination of nuclear binding energies and the associated uncertainties from cross-section measurements. We have illustrated this idea by determining the binding energy of $^{75}$Cu. The extrapolation of cross-sections towards neutron excess can determine the neutron drip-line. We have illustrated this idea with the cases of $^{40}$Mg and $^{39}$Na.

**ACKNOWLEDGEMENTS**

We would like to thank the cyclotron operation staff at Riken for maintaining a high-intensity $^{86}$Kr beam making the cross-section measurements of the long chain of Cu isotopes possible. This work is supported by the National Science Foundation under grants PHY-01-10253, PHY-0606007, INT-0218329, and OISE-0089581.

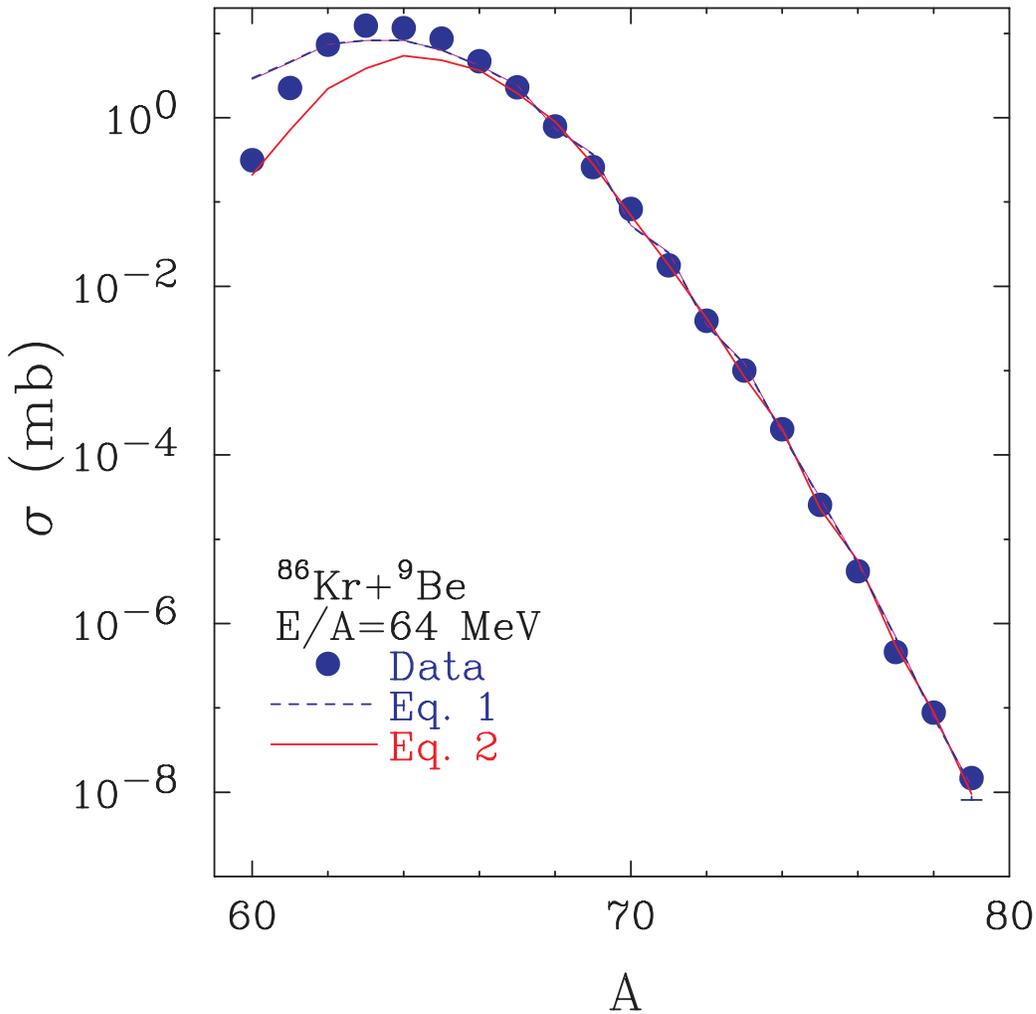

FIG 1: Cross-sections of copper isotopes [8] produced in the projectile fragmentation of $^{86}$Kr+$^9$Be reaction plotted as a function of mass number, A. The dashed line is the best fit from Eq. (1) and solid line is the best fit from the empirical relationship of Eq. (2). The thin solid line which cannot be distinguished from the dashed line is the best fit of Eq. (1) without fitting the isotopes greater than $^{74}$Cu.

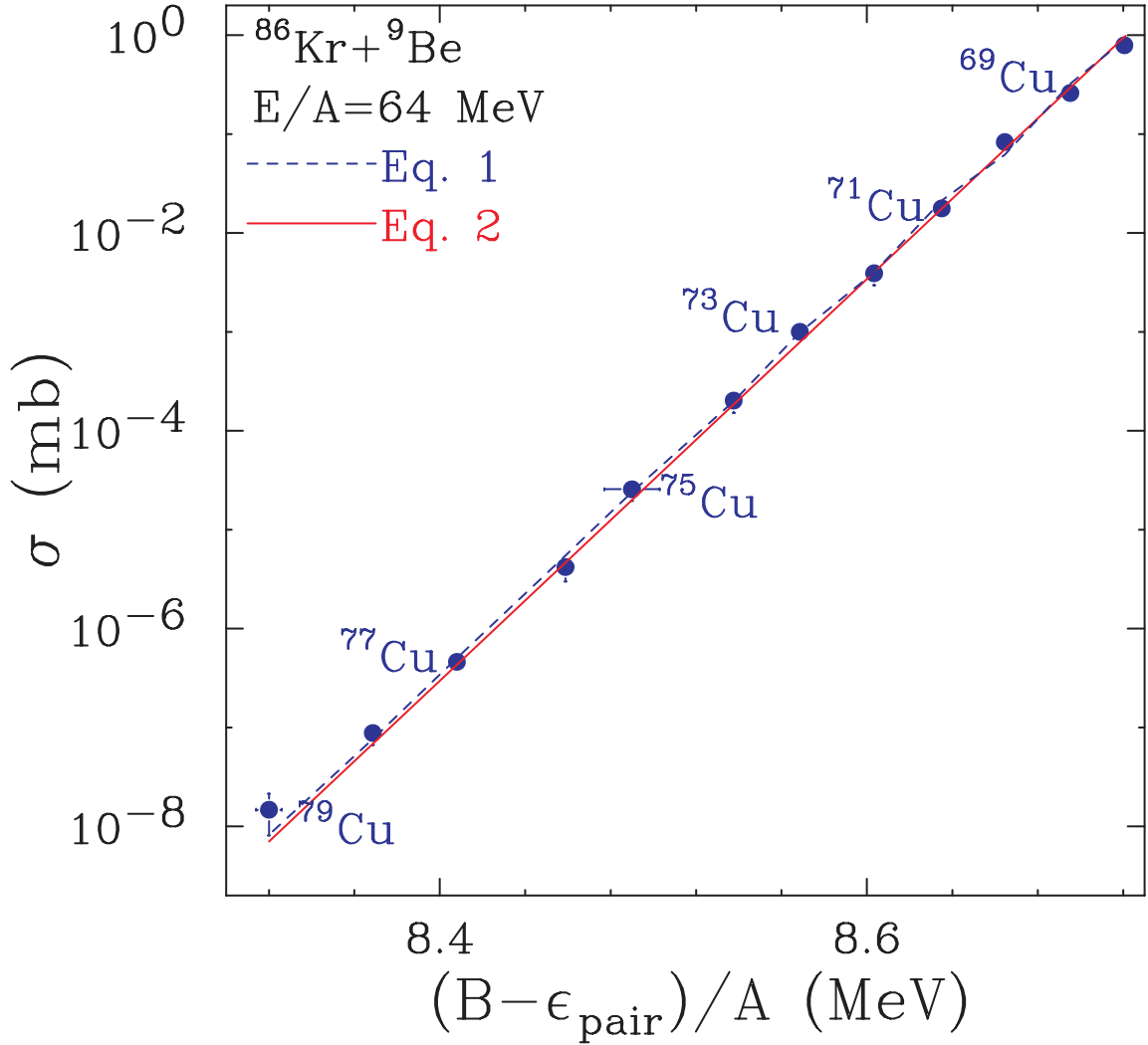

FIG 2: Fragmentation cross-sections of neutron-rich isotopes of copper produced in the projectile fragmentation of $^{86}$Kr+$^{9}$Be reaction, plotted as a function of average binding energy after correcting for the odd-even stagger arising from pairing. The dashed line is the best fit from Eq. (1) and the solid line is the best fit from Eq. (2).

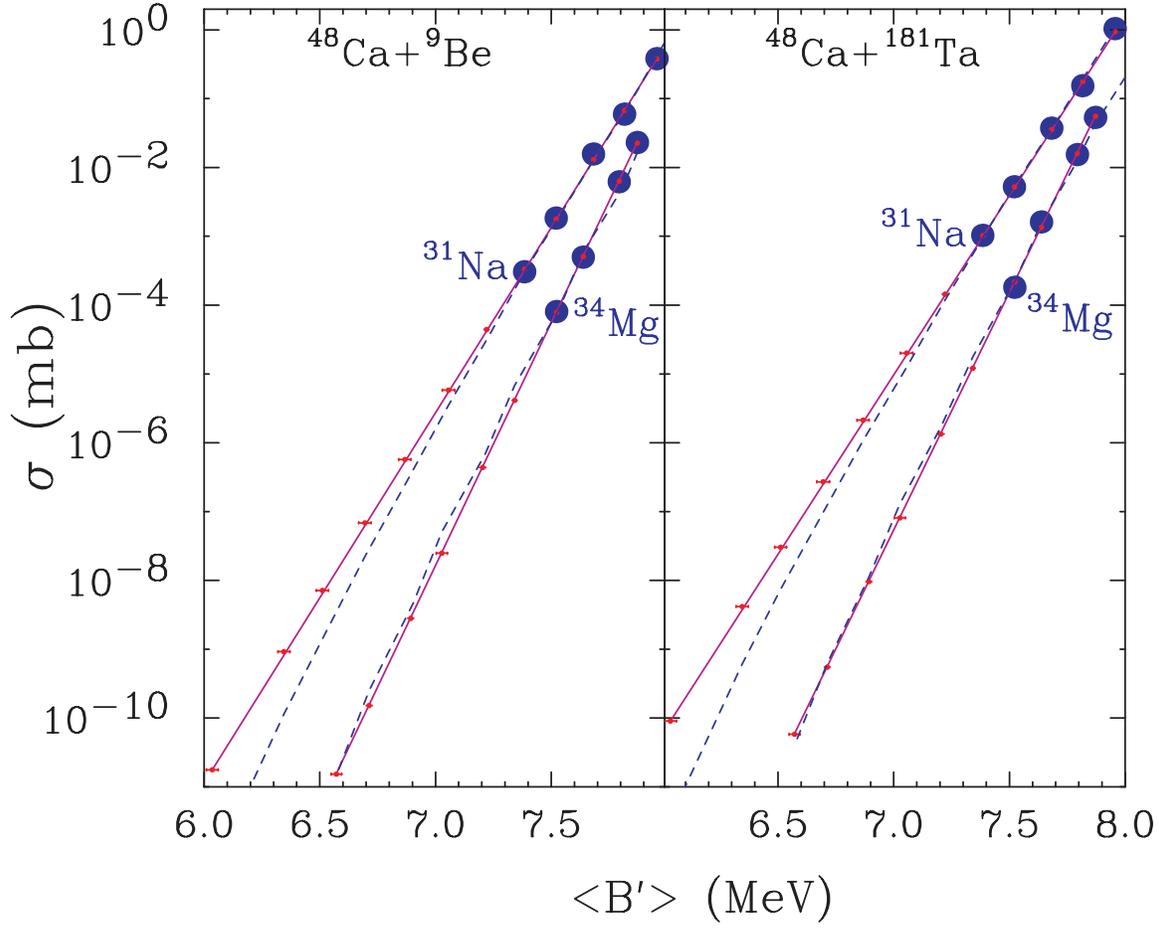

FIG 3: Fragment cross-sections [17] plotted as a function of <B'> for Mg and Na isotopes emitted in $^{48}Ca+^{9}Be$ (left panel) and $^{48}Ca+^{181}Ta$ (right panel) reactions. Dashed lines are best fits from the data using Eq. (1) and solid lines are the best fits from Eq. (2). The solid lines end at the predicted cross-sections of $^{39}Na$ and $^{40}Mg$.